\documentclass[12pt]{article}
\usepackage{epsf}

\usepackage{epsfig,graphics}
\usepackage {graphicx}
\usepackage {epsfig}
\usepackage {psfig}
\usepackage {subfigure}
\usepackage {tabularx} 
\usepackage{rotate}	
\usepackage{slashed}

\newcommand{\bmat}{\left(\begin{array}}
\newcommand{\emat}{\end{array}\right)}

\def\yzero{\smash{\hbox{$y\kern-4pt\raise1pt\hbox{${}^\circ$}$}}}

\def\beq{\begin{equation}}
\def\eeq{\end{equation}}
\def\beqa{\begin{eqnarray}}
\def\eeqa{\end{eqnarray}}

\def\-{\hphantom{-}}

\def\s2{\frac{1}{\sqrt2}}

\def\beq{\begin{equation}}
\def\eeq{\end{equation}}
\def\beqa{\begin{eqnarray}}
\def\eeqa{\end{eqnarray}}

\def\IF{\relax{\rm I\kern-.18em F}}
\def\II{\relax{\rm I\kern-.18em I}}
\def\IP{\relax{\rm I\kern-.18em P}}
\def\IC{\relax\hbox{\kern.25em$\inbar\kern-.3em{\rm C}$}}
\def\IR{\relax{\rm I\kern-.18em R}}

\def\Dsl{\,\raise.15ex\hbox{/}\mkern-13.5mu D} 
\def\IZ{Z\kern-.4em  Z}

\catcode`\@=11   
\newdimen\@rotdimen
\newbox\@rotbox  

\def\@vspec#1{\special{ps:#1}}
\def\@rotstart#1{\@vspec{gsave currentpoint currentpoint translate
   #1 neg exch neg exch translate}}
\def\@rotfinish{\@vspec{currentpoint grestore moveto}}
\def\@rotr#1{\@rotdimen=\ht#1\advance\@rotdimen by\dp#1%
   \hbox to\@rotdimen{\hskip\ht#1\vbox to\wd#1{\@rotstart{90 rotate}%
   \box#1\vss}\hss}\@rotfinish}
\def\@rotl#1{\@rotdimen=\ht#1\advance\@rotdimen by\dp#1%
   \hbox to\@rotdimen{\vbox to\wd#1{\vskip\wd#1\@rotstart{270 rotate}%
   \box#1\vss}\hss}\@rotfinish}%
\def\@rotu#1{\@rotdimen=\ht#1\advance\@rotdimen by\dp#1%
   \hbox to\wd#1{\hskip\wd#1\vbox to\@rotdimen{\vskip\@rotdimen
   \@rotstart{-1 dup scale}\box#1\vss}\hss}\@rotfinish}%
\def\@rotf#1{\hbox to\wd#1{\hskip\wd#1\@rotstart{-1 1 scale}%
   \box#1\hss}\@rotfinish}%
\def\rotate{\@ifnextchar[{\@rotate}{\@rotate[l]}}
\def\@rotate[#1]#2{\setbox\@rotbox=\hbox{#2}\@nameuse{@rot#1}\@rotbox}

\catcode`\@=12

\topmargin
-1.5cm
\textwidth
15.5cm
\textheight
23.5cm
\oddsidemargin
0.7cm
\evensidemargin
0.7cm

\begin{document}

\makeatletter
\@addtoreset{equation}{section}
\makeatother
\renewcommand{\theequation}{\thesection.\arabic{equation}}
\pagestyle{empty}
\rightline{ IFT-UAM/CSIC-12-30}
\vspace{1.5cm}
\begin{center}
\LARGE{From Strings to the LHC}

\vspace{0.5cm}

\large{Les Houches Lectures on String Phenomenology} 
  \\[5mm]
  
  \vspace{2.5cm}
  
 \large{ Luis E. Ib\'a\~nez \\[6mm]}
\small{
 Departamento de F\'{\i}sica Te\'orica 
and Instituto de F\'{\i}sica Te\'orica  UAM-CSIC,\\[-0.3em]
Universidad Aut\'onoma de Madrid,
Cantoblanco, 28049 Madrid, Spain 
\\[8mm]}
\small{\bf Abstract} \\[10mm]
\end{center}
\begin{center}
\begin{minipage}[h]{15.0cm}
These notes are based on lectures given at the Les Houches Summer School in 2011, which was 
centered on the general  topic {\it Theoretical Physics to face the challenge of LHC}.
In these lectures I reviewed a number of topics in the field of String Phenomenology,  focusing on  orientifold/F-theory
models yielding semi-realistic low-energy physics. The emphasis was on the extraction of the low-energy
effective action and the possible test of specific models at LHC. These notes are a brief summary, appropriately updated,  of some of the
main topics covered in the lectures.

\end{minipage}
\end{center}
\newpage
\setcounter{page}{1}
\pagestyle{plain}
\renewcommand{\thefootnote}{\arabic{footnote}}
\setcounter{footnote}{0}


\section{Branes and chirality}

String Theory (ST) is the most serious candidate for a consistent theory of quantum gravity  coupled to matter.
In fact ST actually predicts the very existence of gravity, since a massless spin-2 particle, the graviton 
appears automatically  in the spectrum of closed string theories. String Theory  has also allowed us to 
improve our understanding 
of the origin of the blackhole degrees of freedom and also provides for explicit
realizations of  {\it holography}  through the AdS/CFT correspondance.
Remarkably, ST is not only a theory of quantum gravity 
but incorporates all the essential ingredients of the Standard Model (SM) 
of Particle Physics: gauge interactions, chiral fermions, Yukawa couplings...
It is thus a strong candidate to provide us with a unified theory of all interactions,
including the Standard Model (SM) and gravitation.
\begin{figure}[htbp]
\begin{center}
\includegraphics[angle=90,width=6cm]{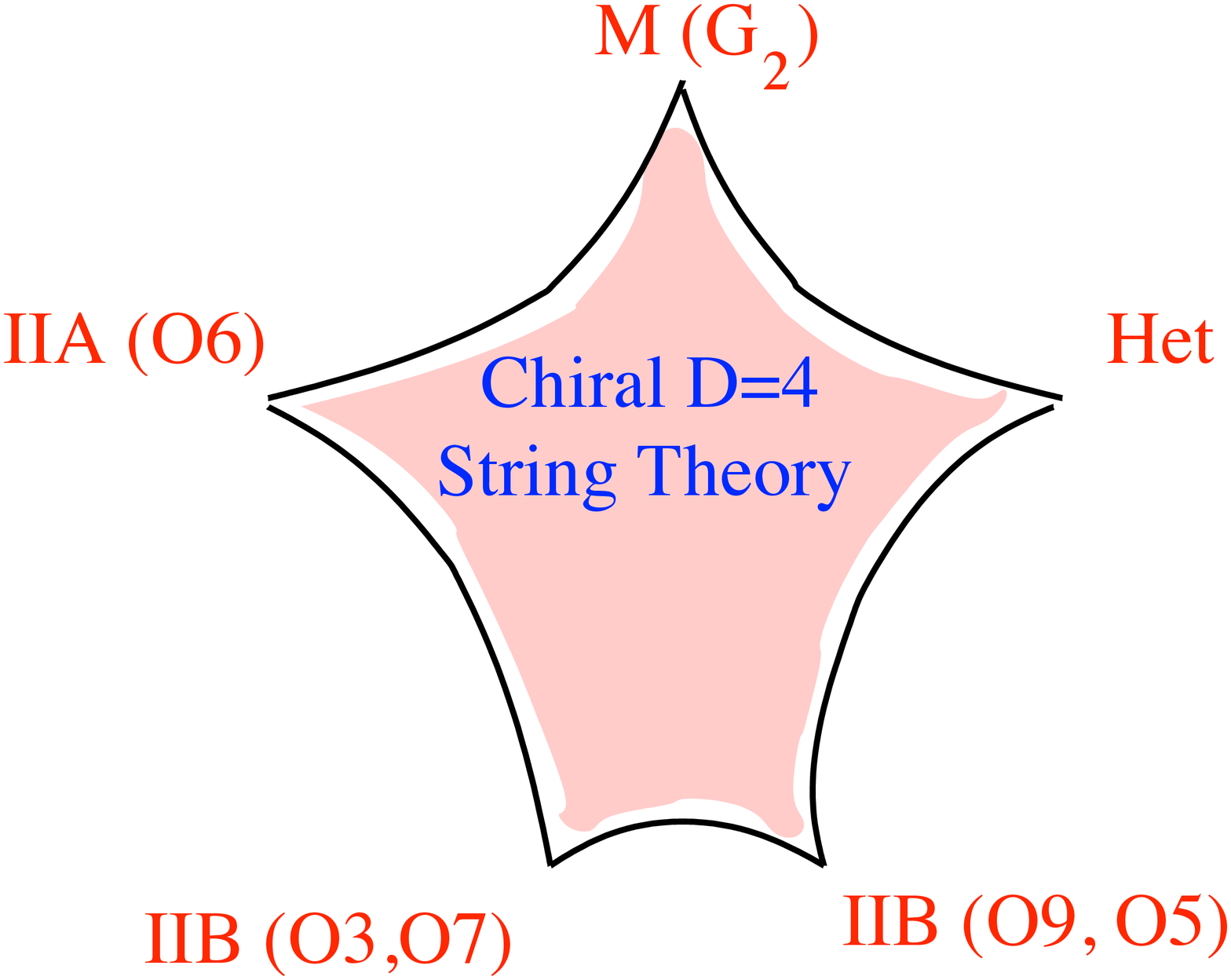}
\end{center}
\caption{The five  large classes of 4d chiral string compactifications.}
\label{pentagono}
\end{figure}
In the last 25 years enormous progress has been obtained in the understanding of the space of 4d 
string vacua \cite{BOOK}.
From the point of view of unification, the main objective is to understand how
the SM may be obtained as a low-energy limit of string theory. We would like to understand how the
SM gauge group, the 3 quark/lepton generations, chirality, Yukawa couplings, CP-violation, neutrino masses,
Higgs sector etc. may appear from an underlying string theory.   The  first step in that direction is 
learning which compactifications  lead to a chiral spectrum of massless fermions at low-energies.
There are essentially five large classes of such chiral 4d string vacua symbolized by the 5
vertices of the {\it pentagon} in fig.(\ref{pentagono}).

These include  three large classes of Type II  {\it orientifolds}  (IIA with  O6 orientifold planes, 
and IIB with O3/O7   or  O9/O5 orientifold planes). In addition there are the well studied heterotic vacua 
in Calabi-Yau (CY) manifolds. Finally there are less studied (and difficult to handle) vacua obtained from the
11d M-theory compactified in manifolds of $G_2$ holonomy. Different dualities connect these different corners so 
the different classes of vacua should be considered as 5 different corners of a single underlying class of
theories.  It is impossible to overview all these different classes of theories so that we will concentrate 
on the case of the  Type II orientifolds whose  potential for the construction of realistic SM-like 
compactifications has been explored in the last 15 years. 

The essential objects in chiral Type II orientifolds are $Dp$-branes, non-perturbative solitonic states
of string theory which extend over  (p+1) space+time dimensions. For our purposes Dp-branes may be considered
as subspaces of the 10d space of Type II string theory in which open strings are allowed to start and end. They are charged
under antisymmetric tensors of the Ramond-Ramond (RR) sector of Type II theory with (p+1) indices. 
Since in Type IIA(IIB) the massless RR tensors have an odd(even) number of indices, there are Dp-branes with p even(odd) 
for Type IIA(IIB) string theory.  We will be interested in Dp-branes large enough to contain the standard Minkowski 
space inside so that the relevant Dp-branes will be D4,D6,D8 in Type IIA and D3,D5,D7,D9 in Type IIB.
In compactified theories
Gauss theorem will force the overall RR charges with respect to these antisymmetric fields to vanish. This leads to
the so called {\it tadpole} cancellation conditions which turn out to also insure cancellation of gauge and gravitational 
anomalies in the theory.

\begin{figure}[htbp]
\begin{center}
\includegraphics[angle=00,width=4cm]{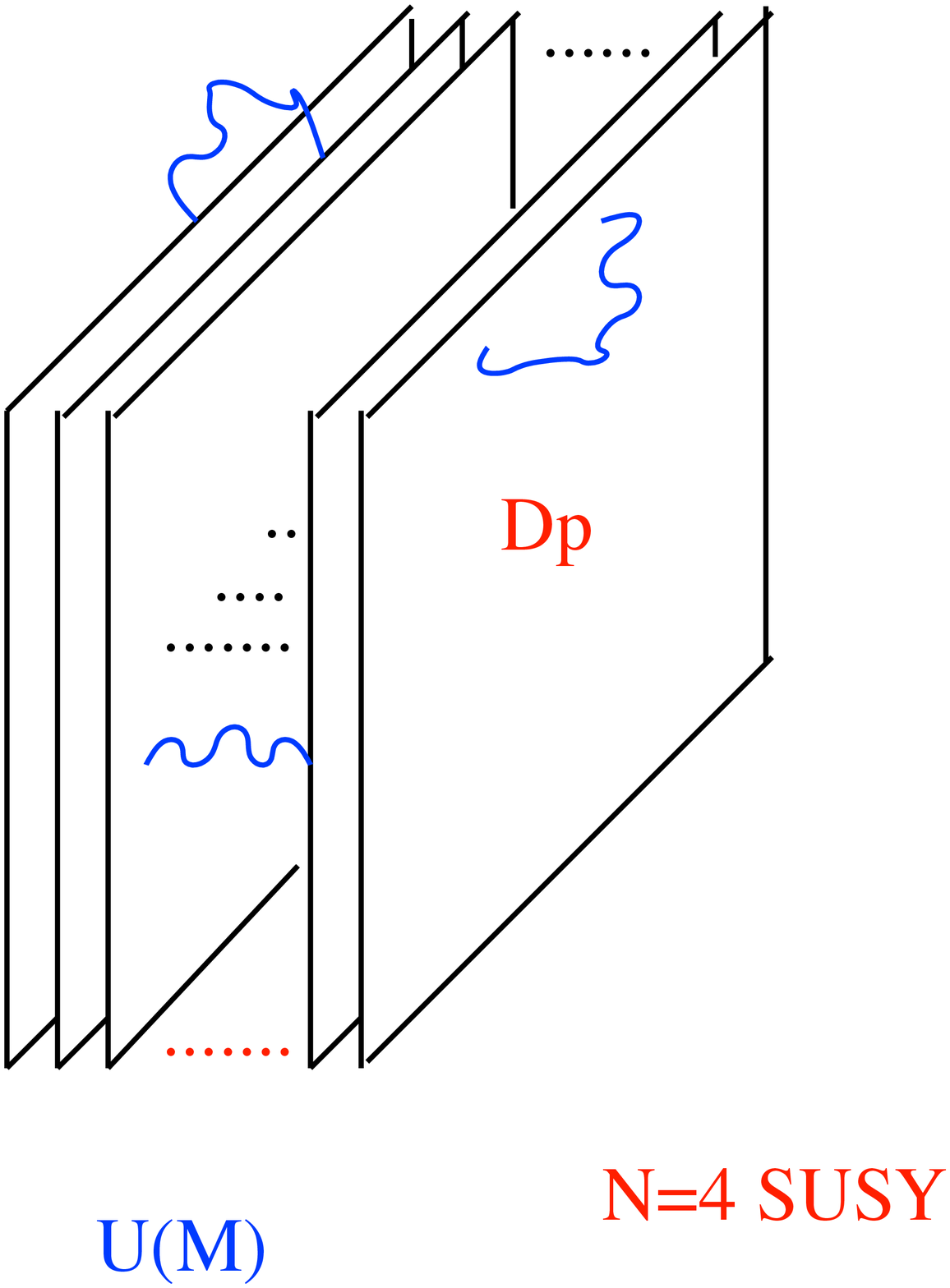}
\end{center}
\caption{Open strings ending on a  stack of M parallell Dp-branes give rise to a U(M), N=4 gauge theory.}
\label{stack}
\end{figure}

In the worldvolume of Dp-branes there live (are localized) gauge and charged matter degrees of freedom. In a single Dp-branes lives
a $U(1)$ gauge boson and M such branes  located in the same place in transverse dimensions contain
an enhaced $U(M)$ gauge symmetry with $N=4$ SUSY in flat space. The corresponding spectrum is obviously non-chiral 
and insufficient to yield realistic physics. In order to obtain chirality additional ingredients must be present.
In the case of Type IIA models with the  six extra dimensions compactified in a Calabi-Yau (CY) manifold,
chiral fermions appear at the intersection of pairs of D6-branes, as we will describe later.  In the case of Type 
IIB models chiral fermions may appear at the worldvolume of D7 or D9 branes in the presence of magnetic fluxes 
in the compact directions. Alternatively, chirality may appear if the geometry is singular, like e.g. the case of 
D3-branes on ${\bf Z_N}$ orbifold singularities.

The other crucial ingredient in perturbative Type II models are  Op {\it orientifolds}.  These are geometrically analogous to 
Dp-branes with the crucial difference that they are not dynamical and do not contain any field degrees of freedom
in their worldvolume. They are however charged under the RR antisymmetric fields and they have also {\it negative}
tension compared to their Dp-branes counterparts. It is precisely these two properties which make useful the
presence of orientifold planes, their negative tension and RR charges may be used to cancel the positive 
contribution of Dp-branes, allowing for the construction of Type II vacua with zero vacuum energy (Minkowski) 
and overall vanishing RR charges in a compact space.

Another important property of Type IIA and IIB vacua is Mirror symmetry. This is a symmetry which exchanges 
IIA and IIB compactifications by exchanging accordingly the underlying CY space by its mirror. For each CY 
manifold one can find a mirror manifold in which the Kahler and complex structure moduli  are exchanged. 
In simple examples  (like tori and orbifolds thereof) one can show that mirror symmetry is a 
particular example of {\it T-duality}.  The action of T-duality in these toroidal/orbifold settings (to be dicussed below) 
is non-trivial and exchanges Neumann and Dirichlet open string boundary conditions. An odd number of T-dualities 
along one-cycles exchanges Type IIA and IIB theories and the dimensionalities of Dp-branes changes accordingly.
Thus e.g. 3 T-dualities on Type IIB D9-branes on $T^6$ change them into $D6$ branes wrapping a 3-cycle in $T^6$. 

The basic rules for D-brane model building are as follows \cite{interev}. One starts with Type II theory compactified on a CY
(in some simple examples one may consider $T^6$ tori or orbifolds). One then consider possible  distributions 
of Dp-branes containing Minkowski space and preserving $N=1$ SUSY in 4d.
  The branes wrap subspaces (cycles) or are located at specific regions inside the CY.  The configuration so far 
has positive energy and  RR charges and is untenable if one wants to obtain Minkowski vacua. To achieve that, 
appropriate Op orientifold planes will be required both to cancel the positive vacuum energy and overall 
RR charges. This will require the construction of a CY orientifold. Finally, the brane distribution is so chosen
that the massless sector resembles as much as possible the SM or the MSSM. If the brane distribution respects the
same $N=1$ SUSY in 4d the theory will be perturbatively stable.

In the above enterprise two approaches are possible:
\begin{itemize}
\item 
{Global models.} One  insists in having a complete globally consistent CY compactification, with 
all RR tadpoles canceling. 
\item
{Local models.} One considers local sets of lower dimensional Dp-branes ($p\leq 7$) which are localized on 
some region of the CY and reproduce the SM or MSSM physics there. One does not care at this stage  about
global aspects of the compactification and assumes that eventually the configuration may be embedded inside a fully
consistent global compact model.
\end{itemize}
The latter is often called the {\it bottom-up} approach \cite{aiqu}, since one first constructs the local (bottom) model 
with the idea that eventually one may embed it in some global model. Note that this philosophy is not applicable to
heterotic or Type I vacua since in those strings the SM fields live in the bulk six dimensions of the CY.

\section{Type II orientifolds: intersections and magnetic fluxes}

\begin{figure}[htbp]
\begin{center}
\includegraphics[angle=00,width=10cm]{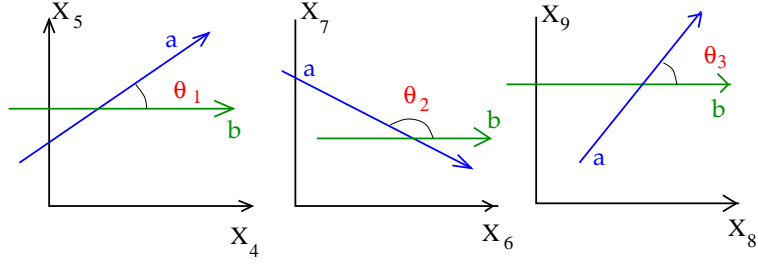}
\end{center}
\caption{Open strings between  D6$_a$-D6$_b$ branes  intersecting  at angles  yield massles chiral fermions. Here $X_5,..,X_9$ are local coordinates 
for the compact space.}
\label{interseccion}
\end{figure}

In Type IIA compactifications in principle we have D4,D6 and D8-branes, big enough to contain Minkowski space {\bf $M_4$}.
They can span  {\bf $M_4$} and wrap respectively 1-, 3- and 5-cycles in the CY. However, since CY manifolds do not have
non-trivial 1- or 5-cycles, in IIA orientifolds only D6-branes are relevant for our purposes.  It is easy to see that a pair of 
intersecting branes,  D6$_a$, D6$_b$ give rise to chiral fermions at their intersection from open strings starting in one and
ending on the other brane (se fig.(\ref{interseccion})).
The mass formula for the fields at an intersection in flat space is given (in bosonized  formulation) by 
\begin{equation}
 M_{ab}^2 
\ =\ 
N_{osc} \ + \frac { ( {r} +  {r_\theta} )^2}{2} \ - \frac {1}{2} \ + \
\sum_{i=1}^3\frac {1}{2} | {\theta_i}|(1-| {\theta_i}|)
\end{equation}
where $r_\theta=(\theta_1,\theta_2,\theta_3,0)$ and
$r$ belongs to the $SO(8)$ lattice ($r_i={\bf Z},{\bf Z}+1/2$ for NS,RR sectors respectively, with $\sum_ir_i=$ odd). 
The reader can check that the state $r+r_\theta=(-\frac {1}{2}+\theta_1,-\frac {1}{2}+\theta_2,
-\frac {1}{2}+\theta_3,+\frac {1}{2})$ is massless for any value of the angles, so there is always a massless fermion
at the intersection. If there are N D6$_a$ and M  D6$_b$ intersecting stacks of  branes the fermion transforms in the bi-fundamental
$(N,{\overline M})$.  There are also three scalars  (e.g. $r+r_\theta=(-1+\theta_1,\theta_2,\theta_3,0)$) with mass$^2$  which
may be positive, zero or negative, depending on the values of the angles. Tachyons are avoided for large ranges of the intersecting angles.
On the other hand for particular choices of the angles there is
a massless scalar, the partner of the chiral fermion, signaling the presence of a $N=1$ SUSY, at least at the local level.

In order to construct 4d models one compactifies Type IIA string theory down to four dimensions on a CY manifold. The resulting
theory has $N=2$ supersymmetry and is not yet suitable for realistic model building. One then constructs an orientifold by moding the theory
by $\Omega{\cal R}$ where $\Omega$ is the worldsheet parity operation and ${\cal R}$ is a ${\bf Z_2}$ antiholomorphic involution on the CY 
with ${\cal R}J=-J$ and ${\cal R}\Omega_3={\overline \Omega }_3$ ($J$ and $\Omega_3$ are the Kahler 2-form and the holomorphic 3-form
characteristic of CY manifolds).  The resulting theory has now $N=1$ SUSY in 4d and the submanifolds left fixed under the ${\cal R}$ operation
are orientifold  O6-planes carrying    $C_7$ RR antisymmetric field charge. To flesh out these process let us consider the simplified 
(yet phenomenologically interesting)  case of a $T^6$ orientifold compactification \cite{intersectionsori}.

\begin{figure}[htbp]
\begin{center}
\includegraphics[angle=00,width=9cm]{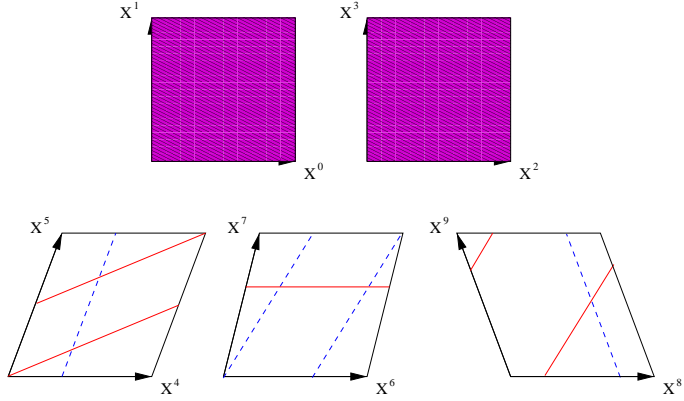}
\end{center}
\caption{D6-branes wrap 1-cycles in each of the 3 $T^2$ and intersect at angles. Chiral bifundamental fermions
are localized at the intersections.}
\label{torito}
\end{figure}

Consider Type IIA string theory compactified in a factorized torus $T^6=T^2\times T^2\times T^2$. D6-branes are assumed to
wrap ${\bf M_4}$ and a 3-cycle which is the direct product of 3 1-cycles, one per $T^2$ (see fig.\ref{torito}).
These cycles are described by integers $(n_a^i,m_a^i), i=1,2,3$  indicating the number of times $n_a^i(m_a^i)$ the D6$_a$ brane wraps
around the  horizontal(vertical)  directions. For each stack of $N_a$ D6$_a$-branes there is a $U(N_a)$ gauge group. Furthermore at the 
intersection of two stacks of  branes D6$_a$, D6$_b$ the exchange of open strings gives rise to massless chiral fermions in bifundamental
$({\bf N_a},{\bf  {\overline {N}}_b})$ representations. Their multiplicity is given by their intersection number
\begin{equation}
I_{ab} \ =\ I_{ab}^1\times I_{ab}^2\times I_{ab}^3\ =\ (n_a^1m_b^1-m_a^1n_b^1)(n_a^2m_b^2-m_a^2n_b^2)(n_a^3m_b^3-m_a^3n_b^3).
\label{internumber}
\end{equation}
which is e.g. $2\times 2\times1=4$  in the example of fig.(\ref{torito}). We now construct an orientifold by moding 
out the theory by the world-sheet operator 
 $\Omega (\tau, \sigma)=(\tau,-\sigma)$ acting on the world-sheet coordinates. Simultaneously we act with a reflection on the
 three coordinates $R(X_i)=-X_i$, $i=5,7,9$. This geometrical reflection leaves invariant the  space  defined by
 $X_5=X_7=X_9=0$ in which the O6 orientifold lives.  In addition the orientifold projection on invariant states may modify the gauge group of the branes 
 if the latter wrap a 3-cycle which is left invariant by the orientifold. Depending on the details of the projection one may get 
 $Sp(N)$ or $O(N)$ groups. On the other hand, if the 3-cycle wrapped by the D6-brane stack  is not invariant, one must add
 in the background extra {\it mirror} D6$^*$ branes siting on the reflected 3-cycle with wrapping numbers
 $(n^i,-m^i)$. Then the configuration is also invariant but the
 gauge group $U(N)$ remains. 

 It is easy to find choices of D6-branes with appropriate wrapping numbers $(n^i,m^i)$ 
yielding a semirealistic chiral massless spectrum. Let as consider  3 stacks  D6$_a$, D6$_b$, D6$_c$ of branes 
on rectangular $T^2$ tori with multiplicities
and wrapping numbers as in table {\ref{tabliguay}} \cite{yuk1}.
\begin{table}[htb] 
\center
\begin{tabular}{|c||c|c|c|}
\hline $N_\alpha $ & $(n_i^1,m_i^1)$ & $(n_i^2,m_i^2)$ & $(n_i^3,m_i^3)$ \\
 \hline\hline $N_a=3+1$ & $(1,0)$ & $(3,1)$ & $(3 , -1)$ \\ $N_b=1$ &
 $(0,1)$ & $ (1,0)$ & $(0,-1)$ \\ $N_c=1$ & $(0,1)$ & $(0,-1)$ &
 $(1,0)$ \\
\hline \end{tabular}
\caption{Wrapping numbers of D6-branes in a MSSM-like configuration.}
\label{tabliguay}
\end{table}
The  D6$_b$ and D6$_c$ branes are assumed to be located at $X_7=0$ and $X_9=0$ respectively
so that the orientifold projection yields an $Sp(1)\simeq SU(2)$ gauge group for both of them. On the other
hand the $3+1$  D6$_a$ branes in the table should be suplemented by their mirrors with wrapping numbers 
flipped as $(n_a^i,m_a^i)\rightarrow (n_a,-m_a^i)$, and the gauge group is $U(3)\times U(1)$. 
The complete gauge group is then $U(3)\times SU(2)\times SU(2)\times U(1)$ but one linear combination of the two
$U(1)$'s is anomalous and becomes massive through a generalized Green-Schwarz mechanism. All in all, one 
obtains the gauge group of the minimal left-right symmetric extension of the MSSM. The reader may check using eq.(\ref{internumber})
that one has three generations of quarks and leptons, with three right-handed neutrinos. Furthermore, if the branes D6$_b$ and D6$_c$ 
sit on top of each other in the first $T^2$, there is one minimal set of Higgs fields.  Chosing  $R_x^2/R_y^2=R_x^3/R_y^3$ for the radii
in the second and third torus one can see that $\theta_2+\theta_3=0$ and there is one unbroken $N=1$ SUSY. 

The above example is a {\it good local model} but it is globally inconsistent.
The reason is that as it stands it gives rise to RR tadpoles, the overall charge with respect o the $C^7$ RR forms does not
vanish as it should in a compact space. It is easy to show that those conditions in this toroidal setting are
\begin{equation}
\sum_aN_an_a^1n_a^2n_a^3\ =\ 16 \ \ ;\ \ \sum_a N_an_a^1m_a^2m_a^3\ =\ 0 \ \ (+\ permutations)
\label{tadpoles}
\end{equation}
and plugging the wrapping numbers of the table one observes they are not obeyed. It is however easy to constract 
a ${\bf Z_2}\times {\bf Z_2}$ orbifold variation  of this model with some additional D6-branes and orientifold planes which is supersymmetric and
obeys the corresponding tadpole conditions \cite{z2z2}.

This model is remarkably simple and its chiral sector gets quite close to a phenomenologically interesting model,
the L-R extension of the MSSM. Still has the shortcoming that, like most  toroidal/orbifold models, the massless spectrum includes
additional adjoint chiral multiplets of the SM gauge group. The vev of these adjoints parametrize the freedom to translate in parallel the
positions of the branes in any of these models. The latter is a characteristic of toroidal compactifications and is in general absent in
more general CY orientifolds. 

A second class of interesting Type II compactifications is Type IIB orientifolds. Now the internal
orientifold geometric involution acts like $RJ=J$ , $R\Omega_3=-\Omega_3$.
In the toroidal setting they may be obtained as T-duals of Type IIA  intersecting brane models. Indeed, upon an odd number of T-dualities
along the 6 circles  in the $T^6$ a D6-brane may transform into a D9,D7,D5 or D3-brane, depending on the particular T-duality transformation.
If the original D6 brane is rotated with respect to the orientifold plane the resulting IIB Dp-branes will in general contain
a magnetic flux turned on in their worldvolume.
Indeed,  higher dimensional 
Type IIB branes  in SUSY configurations (unlike the D6-branes in IIA)  may contain magnetic flux backgrounds. They in turn
 induce lower dimensional Dp-brane charge and also  chirality. Let us consider \cite{magnetized} 
 the case of $N_a$ D9-branes wrapped $m_a^i$
times on the i-th $T^2$ and with $n_a^i$ units of $U(1)_a$ quantized magnetic flux:
\begin{equation}
m_a^i\frac {1}{2\pi}\int_{T_i^2}\ F_a^i\ =\ n_a^i \ .
\end{equation}
Interestingly enough the   $(n_a^i,m_a^i)$ D6 wrapping numbers are mapped under T-duality into the magnetic integers defined above.
In addition the relative angle $\theta_{ab}^i$ of D6$_a$-D6$_b$ branes  in the i-th torus is mapped into the difference
\begin{equation}
\theta_{ab}^i\ =\ arctg(F_b^i)\ -\ arctg(F_a^i) \ \ \ ,\ \ \ F_a^i\ =\ \frac {n_a^i}{m_a^iR_{xi}R_{yi}} \ .
\end{equation}
In the presence of a magnetic flux $F$ in a IIB brane wrapping $T^2$ the open string boundary conditions get mofified as
\begin{equation}
\partial_\sigma X \ -\ F\partial_\tau Y \ =\ 0\ \ ;\ \  \partial_\sigma Y \ +\ F\partial_\tau X \ =\ 0 .
\end{equation}
In particular varying $F$ one interpolates between Neumann and Dirichlet boundary conditions and e.g. 
at formally infinite flux they are purely Dirichlet.  Thus adding  fluxes on a higher dimensional brane induces
RR charge corresponding to lower dimensional branes. For example,  D9 branes with  flux numbers
$(1,0)(n_a^2,m_a^2)(n_a^3,m_a^3)$  are equivalent to D7$^1$ branes which are localized on the first $T^2$
and wrap the remaining $T^2\times T^2$. On the other hand D9 branes with flux numbers $(1,0)(1,0)(1,0)$
(formally infinite flux in the three $T^2$'s) is equivalent to D3-branes. Note in particular that the semirealistic 
model with intersecting D6-branes as in the table above are mapped into a set of three stacks of 
D7$_a^1$, D7$_b^2$, D7$_c^3$ which overlap pairwise on a $T^2$.  Chirality arises in this Type IIB mirror
from the mismatch between L- and R-handed fermions induced by the finite flux in the second and third tori.

This view of the orientifolds in terms of Type IIB D7-branes overlapping on dimension 2 spaces ($T^2$ in the
toroidal example) is particularly interesting because it admits a straightforward generalization to Type IIB
CY orientifolds, at least in the large compact volume approximation in which  Kaluza-Klein field theory techniques
are available. 
On the contrary, the mirror class of models of Type IIA orientifolds with intersecting D6-branes is 
more difficult to generalize to curved CY spaces since the mathematical definition of BPS D6-branes in curved 
space  (wrapping so called Special Lagrangian 3-cycles) is more difficult to analyze.  
A further argument to concentrate on Type  IIB orientifolds with D7/D3 branes is that in the last 10 years 
we have learnt a great deal about how the addition of IIB closed string antisymmetric field fluxes can 
fix most or all the moduli. The equivalent analysis for Type IIA or heterotic vacua is at present 
far less developed.

One generic problem in both IIA and IIB cases is the {\it top quark problem} in models with a unified
gauge symmetry like $SU(5)$. The point is that in perturbative orientifolds the GUT symmetry is actually
$U(5)$ and the quantum numbers of a GUT generation are ${\bf {\bar 5}_{-1}}+{\bf 10_{2}}$, with
Higgs multiplets $ {\bf 5_1}+{\bf {\bar 5}_{-1}}$. It is then clear that  D-quark/lepton Yukawas are allowed by the
$U(1)$ symmetry but the U-quark couplings from ${\bf 10_{2}}{\bf 10_{2}}{\bf 5_1}$ are perturbatively forbidden.
This $U(1)$ symmetry is in fact anomalous and massive but still remains as a perturbative global symmetry in
the effective action. Instanton effects may violate it but one expects the corresponding non-perturbative contributions 
to be small and be relevant at most only for the lightest generations, not the top quark. Thus insisting in unification
of SM groups in perturbative orientifolds gives rise to a {\it top quark problem}.

\section{Local F-theory GUT's}

F-theory \cite{ftheory} may be considered as   a geometric  non-perturbative formulation of Type IIB orientifolds. 
From the model building point of view its interest is twoflod: 1) It provides a solution to the 
{\it top quark problem} of perturbative Type II orientifolds with a GUT symmetry and 2) Moduli fixing
induced by closed string antisymmetric fluxes is relatively well understood.  In loose terms one could say 
that it combines advantages from the heterotic and Type IIB vacua. 

\begin{figure}[htbp]
\begin{center}
\includegraphics[angle=00,width=6cm]{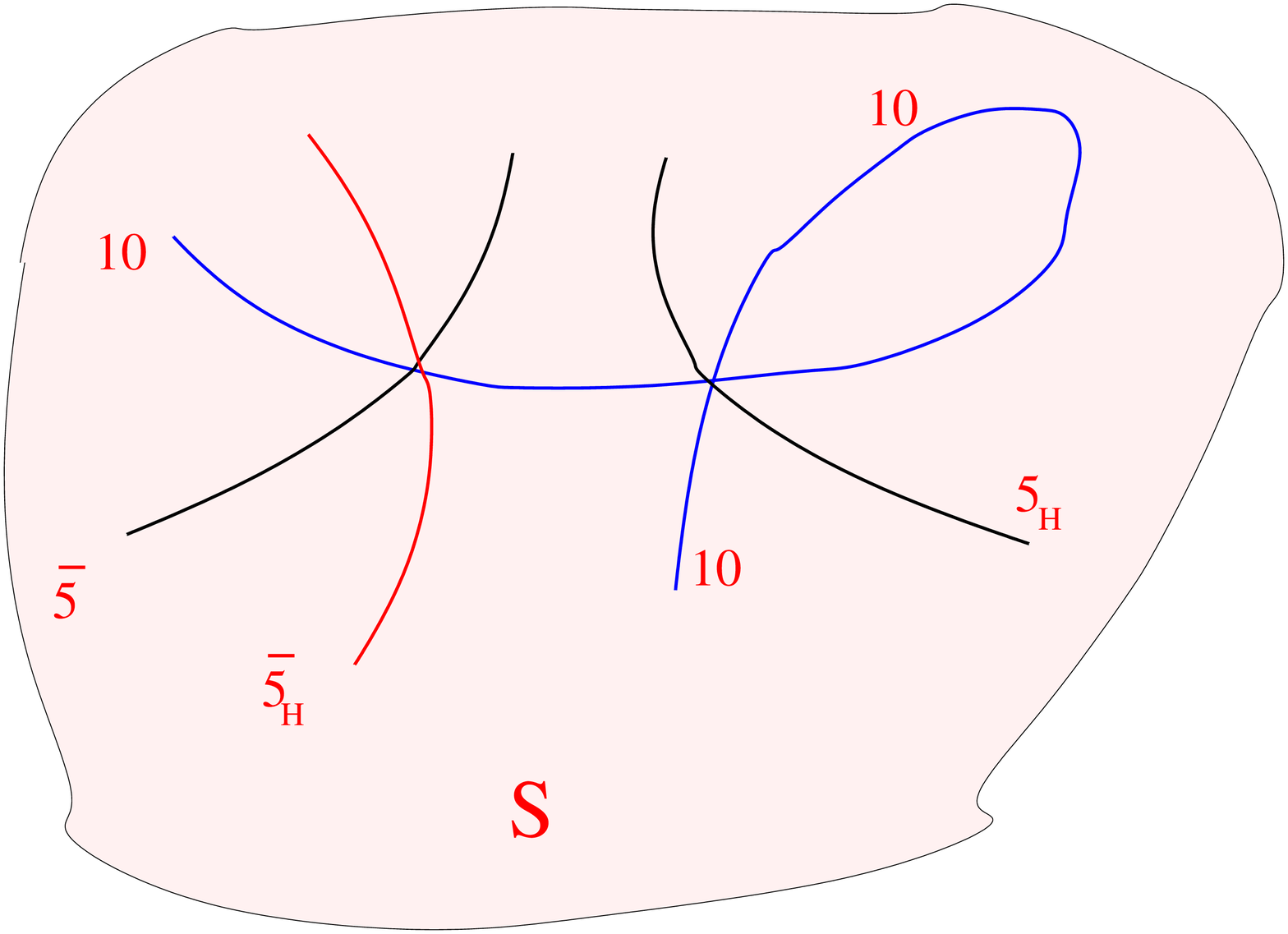}
\end{center}
\caption{Matter curves generically intersect at points of further enhanced symmetry at which
the $10\times {\bar 5}\times {\bar 5}_H$ and $10\times 10\times {\bar 5}_H$ Yukawa couplings localize.}
\label{yukftheory}
\end{figure}

An important massless field of 10d Type IIB string theory is the complexified dilaton field
$\tau = e^{-\phi}+i C_0$. The dilaton $\phi$ controls the perturbative loop expansion and 
$C_0$ is a RR scalar. The 10d theory has a   $SL(2,{\bf Z})$ symmetry under which $\tau$
as the modular parameter. The symmetry is generated by the transformations 
$\tau \rightarrow 1/\tau$ and $\tau\rightarrow \tau + i$ and is clearly non-perturbative 
(e.g. it exchanges strong and weak coupling by inverting the dilaton). F-theory provides a
geometrization of this symmetry by adding two (auxiliary) extra dimensions with  $T^2$ geometry
and identifying the complex structure of this $T^2$ with the Type IIB $\tau$ field. 
The resulting geometric construction is 12-dimensional and one obtains $N=1$ 4d vacua by
compactifying the theory on a complex 4-fold  CY $X_4$ which is an {\it eliptic fibration} over a 6-dimensional
base $B_3$, i.e., locally one has $X_4\simeq T^2\times B_3$. 
The theory contains 7-branes which appear at points in the base $B_3$ at which the fibration
becomes singular, corresponding to 4-cycles wrapped by the 7-brane. As in the case of perturbative 
D7-branes, there is a gauge group associated to these branes. However, unlike the perturbative case, 
the possible gauge groups include the exceptional ones $E_6,E_7,E_8$.  This is an important property
since, as we will see momentarily, allows for the existence of an $SU(5)$ GUT symmetry with a large
top Yukawa.
\begin{figure}[htbp]
\begin{center}
\includegraphics[angle=270,width=9cm]{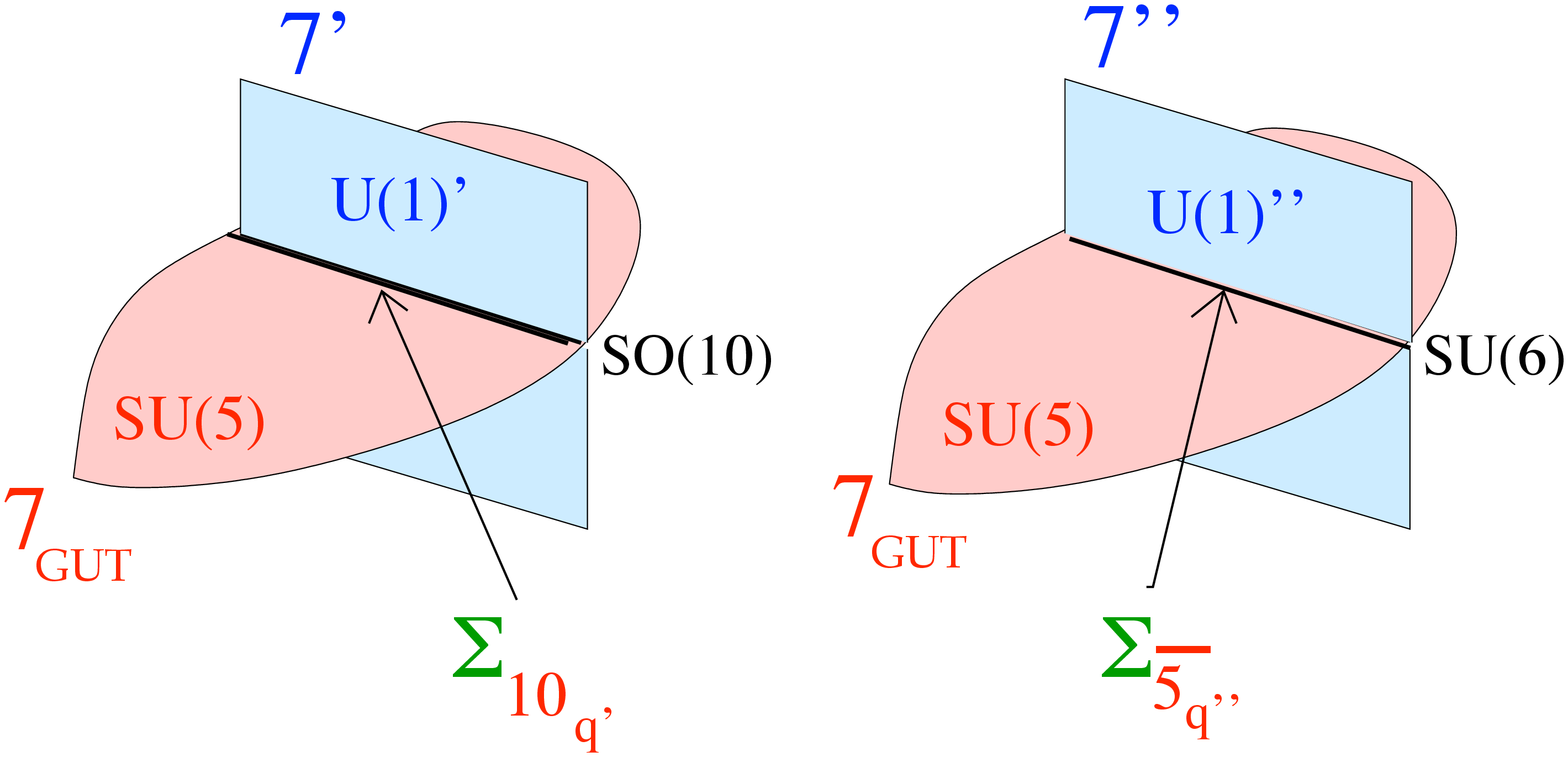}
\end{center}
\caption{The $SU(5)$ matter fields live at matter curves corresponding to the intersection of the
bulk $SU(5)$ brane with $U(1)$ branes. At the matter curves the symmetry is enhanced to
$SU(6)$ and  $SO(10)$ respectively for the multiplets ${\bf 10}$ and ${\bf {\bar 5}}$.}
\label{unfolding}
\end{figure}
A particularly interesting type of F-theory constructions are those involving a GUT symmetry like $SU(5)$, termed F-theory GUT's
\cite{ftheoryGUTs}. 
This is motivated by the apparent unification of coupling constants in the MSSM.
Such  constructions are  a non-perturbative generalization of the Type IIB models with intersecting
(and magnetized) D7 branes that we discussed in previous section.  There is a 7-brane wrapping a 4-cycle 
$S$ inside $B_3$ yielding an $SU(5)$ gauge symmetry.  As in the bottom-up approach mentioned above,
one can decouple the local dynamics associated to the $SU(5)$ brane from the global aspects of the
$B_3$ compact space.
Chiral matter again appears 
at the intersection of pairs of 7-branes, {\it matter curves} in the F-theory language,  corresponding to an 
enhanced degree of the singularity.  These 7-branes are however non-perturbative and cannot simply be described 
in terms of perturbative open strings. A visual intuition of the arisal of matter fields in an
$SU(5)$ F-theory GUT is shown in  fig.(\ref{unfolding}). At the matter curves the symmetry is locally enhanced to
$SU(6)$ or $SO(10)$. Recalling the adjoint branchings
\begin{eqnarray}
SU(6)\ & \longrightarrow & \ SU(5)\times U(1) \\ \nonumber
{\bf 35}  & \longrightarrow &\ {\bf 24}_0 \ +\ {\bf 1}_0 \ +\ [ {\bf 5}_1 \ +\ c.c.] \\ \nonumber
SO(10) \ & \longrightarrow & \ SU(5)\times U(1)' \\ \nonumber
{\bf 45}  & \longrightarrow &\ {\bf 24}_0 \ +\ {\bf 1}_0 \ +\ [ {\bf 10}_4 \ +\ c.c.] \\
\end{eqnarray}
one sees that in the matter curve associated to the 5-plet the symmetry is enhanced to $SU(6)$ whereas
in the one related to the 10-plets the symmetry is enhanced to $SO(10)$. Like in the perturbative magnetized IIB
orientifolds, in order to get chiral fermions there must be in general non-vanishing fluxes along the $U(1)$ and
$U(1)$' symmetries.  
A third matter curve with an enhanced $SU(6)$' symmetry is also required to obtain 
Higgs 5-plets. Yukawa couplings appear at the intersection of the Higgs matter curve with the fermion matter curves, as
ilustrated in fig.(\ref{yukftheory}).
At the intersection point the symmetry is further enhanced to $SO(12)$ in the case of the 
$10\times {\bar 5}\times {\bar 5}_H$ couplings and to $E_6$ in the case of the U-quark couplings. 
One may now understand why there are U-quark Yuyawa couplings in F-theory by looking at the 
branching of $E_6$ adjoint  into $SU(5)\times U(1)\times U(1)$', 
\begin{eqnarray}
E_6 \  & \longrightarrow & \ SU(5)\times U(1)\times U(1)' \\ \nonumber
{\bf 78}\  & \longrightarrow & \  { Adjoints} \ +\  [({\bf 10},-1,-3)+({\bf 10},4,0)+({\bf 5},-3,3)+({\bf 1},5,3)+h.c. ]\ .
\label{esbranching}
\end{eqnarray}
We now see that one can form a $10\times 10\times 5$ coupling which is indeed allowed by the 
$U(1)$ symmetries. We will come back to the issue of Yukawa couplings in F-theory local GUT's
in the next section.

To make the final contact with SM physics the $SU(5)$ symmetry must be broken down to
$SU(3)\times SU(2)\times U(1)$. In these constructions there are no massles adjoints to
make that breaking and discrete Wilson lines are also not available. Still one can make such breaking by the addition of an 
additional  flux $F_Y$ along the hypercharge direction of the $SU(5)$, which has the same symmetry breaking effect
as an adjoint Higgs. Interestingly enough, this hypercharge flux may also be used to obtain
doublet-triplet splitting of the Higgs multiplets ${\bf 5}_H+{\bf {\bar 5}}_H$.

\section{The effective low energy action}

To make contact with the low-energy physics we need to have information about the
effective low-energy action remaining at scales well below the string scale. Here we will
concentrate on  the case of field theories with $N=1$ supersymmetry, which is assumed to
be later broken at scales of order the Electro-Weak (EW) scale. In this case the action is 
determined by Kahler potential $K$, gauge kinetic functions $f_a$ and the superpotential $W$,
that we will discuss in turn. For definiteness we will concentrate also in the case of the
effective action for Type IIB orientifolds, whose general features are also expected to 
apply for the F-theory case.

In the massless sector of a $N=1$ compactification there are charged fields from the open string sector 
(to be identified with the SM fields)  and closed string fields giving rise to singlet chiral multiplets, the moduli.
Among the latter there is the complex dilaton $S=e^{-\phi}+iC_0$ which is just the dimensional reduction 
of the complex dilaton $\tau$ mentioned above. In addition there are $h_{11}$ Kahler moduli $T^i$ and 
$h_{21}^-$ complex structure moduli  $U^j$ (the minus means number of (2,1)-forms odd under the orientifold
projection). The Kahler moduli parametrize the volume of the manifold and also of all the 4-cycles $\Sigma_4^{(i)}$  of the specific CY.
The complex structure fields $U^j$ on the other hand parametrize the deformations of the CY
manifolds and are associated to the 3-cycles $\Sigma_3^{(j)}$ in the CY. Specifically one has \cite{louis} 
(in the simplest $h_{11}^-=0$ case)
\begin{equation}
T^i\ =\ e^{-\phi}Vol(\Sigma_4^{(i)})\ +\ i C_4^{(i)} \ \ ; \ \ U^j\ =\ \int_{\Sigma_3^{(j)}}\ \Omega_3
\label{kahlermodulus}
\end{equation}
Here $C_4^{(i)}$ are 4d  zero modes of the RR 4-form $C_4$ on the 4-cycles. The $N=1$ supergravity Kahler potential
associated to the moduli in Type IIB orientifold compactifications may be written as \cite{louis}
\begin{equation}
K_{IIB} \ =\ -log(S+S^*)\ -\ 2log(e^{-3\phi/2}Vol(CY)) \ -\ log(-i\int \Omega_3\wedge {\overline {\Omega}}_3),
\end{equation}
where $Vol(CY)$ is the volume of the CY manifold. In the toroidal case with rectangular $T^6=T^2\times T^2\times T^2$ 
the Kahler potential takes the simple form
\begin{equation}
K_{IIB} \ =\ -log(S+S^*)\ -\ \sum_{i=1}^3log(U_i+U_i^*)\ -\ 
\sum_{i=1}^3log(T_i+T_i^*),
\end{equation}
where $T_i=e^{-\phi}R_x^jR_y^jR_x^kR_y^k-iC_4$, with $i\not=j\not=k\not=i$ and $U_i=R_y^i/R_x^i$.
This is the familiar no-scale structure which also appears in heterotic $N=1$ vacua. 

Concerning the action for the charged matter fields $\Phi_a$ on the 7-branes, the corresponding Kahler metrics, gauge
kinetic functions and superpotential are themselves functions of the moduli.  One can write for the general form of the
supergravity Kahler potential an  expression (to leading order in a matter field expansion)
\begin{equation}
K(M,M^*,\Phi_a,\Phi_a^*) =K_{IIB}(M,M^*) + \sum_{ab} K_{ab}(M,M^*) \Phi_a\Phi_b^*
+log|W(M) + W_{Y}(M,\Phi_a)|^2\ 
\label{kaleron}
\end{equation}
where $M$ collectively denotes the moduli $S,T^i,U^j$, $W(M)$ is the superpotential of the moduli and
$W_Y(M,\Phi_a)$ the Yukawa coupling superpotential of the SM fields. We have already discussed the first term in eq.(\ref{kaleron})
and we will discuss the rest of the terms in what follows.

\subsection{The Kahler metrics}
Equation (\ref{kaleron}) includes the kinetic term for the matter fields which is controlled by the 
Kahler metrics $K_{ab}$, which is a function of the moduli. This dependence on the moduli is
dictated by the geometric origin of the field.  It  has been computed 
at the classical level for some simple cases (mostly toroidal/orbifold orientifolds) either by 
dimensional redaction from the underlying 10d theory or using explicit string correlators.
We are particularly interested in the Kahler metrics of fields living at intersecting 7-branes,
since those are the ones which are associated to the MSSM fields in semi-realistic IIB or
F-theory compactifications. In the case of type IIB toroidal/orbifold orientifolds the matter fields associated to a 
pair of intersecting D7$^i$-D7$^j$ branes has a metric (neglecting magnetic fluxes for the moment) \cite{rigolin}
\begin{equation}
K_{ab}^{ij}\ =\ \delta_{ab} \ \frac {1}{u_i^{1/2}u_j^{1/2}t_k^{1/2}s^{1/2}}\ \ ,\ \ i\not=j\not=k\not=i
\label{metrictori}
\end{equation}
where $t_i=(T_i+T_i^*), u_i=(U_i+U_i^*)$ and $s=(S+S^*)$.  
We thus see that the metrics of matter fields at untersections scale like $K_{ab}\simeq t^{-1/2}$ with the
Kahler moduli.

\begin{figure}[htbp]
\begin{center}
\includegraphics[angle=00,width=4cm]{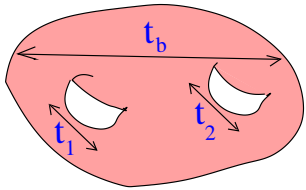}
\end{center}
\caption{ CY manifold with a {\it swiss cheese} structure.}
\label{swiss}
\end{figure}

Toroidal orientifolds/orbifolds, however,   are very special in some ways.
We would rather  like to see to what extent this type of Kahler metrics generalizes to more general IIB CY orientifolds. In particular D7-branes wrap 4-tori whose volumes are directly related to the overall volume of the compact manifold. One would rather like to obtain information
about the Kahler metric when the 7-branes wrap a local 4-cycle whose volume is not directly connected to the overall volume of
the CY. An example of this is provided by the {\it swiss cheese} type of compactifications discussed in ref.\cite{fernando}.
   In this more general setting one assumes that the 
SM fields are localized at D7-branes wrapping {\it small} cycles in a CY whose overall volume is controlled by a large modulus $t_b$ 
(see fig.(\ref{swiss})) so the $Vol[CY]= t_b^{3/2}-h(t_i)$, where h is a homogeneous function of the {\it small} Kahler moduli
$t_i$ of degree $3/2$. The simplest example of this is provided by the CY manifold ${\bf P}^4_{[1,1,1,6,9]}$  which has only two Kahler moduli
$t_b,t$ with a  Kahler potential of the
form
\begin{equation}
K _{IIB} \ =\ -2log(t_b^{3/2}\ -\ t^{3/2}) \ .
\label{kahlerswiss}
\end{equation}
Here  we will assume $t_b\gg t$ and take both large so that the supergravity approximation is still valid. In the F-theory context
the analogue of these
moduli $t,t_b$ would correspond to the size of the 4-fold $S$ and the 6-fold $B_3$ respectively. 
Focusing only in the Kahler moduli dependence of the metrics, one   can write a large volume ansatz  for the Kahler metrics of
charged matter fields at the intersections \cite{cremades}
\begin{equation}
K_\alpha \ =\ \frac {t^{(1-\xi_\alpha )}}{t_b},
\label{metricswiss}
\end{equation}
with $\xi_\alpha$ to be fixed. One can compute  $\xi_\alpha$ by studying the behavior 
with respect to a scaling of $t$ in the effective action. 
In particular in $N=1$ supergravity the physical (i.e. with normalized kinetic terms)  Yukawa coupling 
${\hat Y}_{\alpha \beta \gamma}$  among three chiral fields are related to the holomorphic 
Yukawa coupling  $Y ^{(0)}_{\alpha \beta \gamma}$ by
\begin{equation}
{\hat Y}_{\alpha \beta \gamma}\ = \ e^{K/2} \frac {Y ^{(0)}_{\alpha \beta \gamma}}{(K_\alpha K_\beta K\gamma)^{1/2}} \ .
\end{equation}
On the other hand it is well known that the perturbative holomorphic Yukawa couplings in Type IIB string theory are
independent of Kahler moduli. Then using eqs. (\ref{kahlerswiss}),(\ref{metricswiss}) one finds  a scaling of the physical
Yukawa 
\begin{equation}
{\hat Y}_{\alpha \beta \gamma}\ \simeq \ t^{(\xi_\alpha +\xi_\beta +\xi_\gamma -3)/2} \ .
\label{scaling}
\end{equation}
The dependence on $t_b$ drops at leading order in $t/t_b$, as expected for a model whose physics is essentially
localized on the 4-cycle parametrized by $t$.  On the other hand one can alternatively compute the scaling behavior 
of the physical Yukawa in terms of its computation as an overlap integral of the respective wave functions in $S$
(see below) so that
\begin{equation}
{\hat Y}_{\alpha \beta \gamma}\ \simeq \ \int\ \Psi_\alpha \Psi_\beta \Psi_\gamma  \ \ ,\ \
\int|\Psi_\alpha|^2  = 1.
\end{equation}
For fields localized at intersecting branes the normalization  integrals  above are essentially 2-dimensional  so that the wave functions should scale like 
like $t^{-1/4}$. On the other hand the overlap integral for the Yukawa is essentially point-like so that it scales like ${\hat Y}\simeq t^{-3/4}$. 
Comparing this to eq.(\ref{scaling}) one concludes that all $\xi_\alpha =1/2$ and hence the matter metrics of fields at intersecting branes
have a metric with a local kahler modulus dependence of the form
\begin{equation}
K_\alpha \ =\ \frac {t^{1/2}}{t_b} \ .
\label{metricswiss2}
\end{equation}
Note that setting $t_b\simeq t$ reproduces the Kahler modulus dependence $t^{-1/2}$ of toroidal models eq.(\ref{metrictori}).

\subsection{The gauge kinetic function}

The gauge kinetic function for the gauge group living on the D7-worldvolume may be extracted by 
expanding the Dirac-Born-Infeld (DBI) action of the D7-brane to second order in the gauge field strength $F_a$.
One obtains the general expression \cite{BOOK}
\begin{eqnarray}
f_a^{{\rm D}7} = \frac { (\alpha ')^{-2}} {(2\pi )^5} \, \left(  \,e^{-\phi }\!
\int_{\Sigma_4^a} {\rm Re} \,\left(
 e^{J+i2\pi \alpha ' { F}_a} \right) +i\int_{\Sigma_4^a} \sum_k  C_{2k}\, e^{2\pi \alpha ' { F}_a}  \, \right),
\quad
\label{ffunctionD7}
\end{eqnarray}
where $J$ is the Kahler 2-form and the second piece performs a formal sum over all RR $C_{2k}$ forms contributing to the integral.
Upon expanding the exponential, the first term produces the volume of 4-fold $\Sigma _4^a$ wrapped by the D7  and the second term is
proportional to $C_4$. Taking into account eq.(\ref{kahlermodulus}) one sees that the gauge kinetic function is proportional
to the  Kahler modulus  $T_a$, i.e.
\begin{equation}
f_a \ =\ T_a \ .
\label{gaugekin}
\end{equation}
The subsequents terms in the expansion
describe contributions from the worldvolume gauge magnetic  flux which will be subleading for large volume $t_a=ReT_a$.
In particular, since $\int_{\Sigma_2} F_a=n$, with $n=$ integer for quantized gauge fluxes, one can estimate the flux density as
$F_a\simeq n(ReS/ReT_a)^{1/2}$. The leading flux correction to the gauge kinetic function has then  the form
\begin{equation}
Ref_a \simeq \ = ReT_a(1+|F_a|^2) \ \simeq \ ReT_a(1 + n^2Re(S)/Re(T_a))
\label{fluxcorrgauge}
\end{equation}
which indeed will be subleading in the large $t_a$ limit.

\subsection{The superpotential}

As indicated in eq.(\ref{kaleron}) there will be a superpotential for the moduli $W(M)$ and 
a second superpotential  $W(M,\Phi_\alpha )$ involving (moduli dependent) Yukawa couplings.
In the absence of closed string antisymmetric fluxes the perturbative superpotential of the 
moduli  vanishes, $W_{pert}(M)=0$.  However in the presence of Type IIB NS(RR) 3-form fluxes  $H_3(F_3)$ 
there is an induced effective superpotential involving the complex dilaton $S$ and the complex structure 
moduli $U^j$ given by \cite{gvw}
\begin{equation}
W_{flux} (S,U^j)\ =\ \int_{CY}\ (F_3\ -\ iSH_3)\wedge \Omega_3
\label{gvw}
\end{equation}
where $\Omega_3$ is the CY holomorphic 3-form which may be expanded in terms of the 
complex structure moduli $U^j$ in the CY. This superpotential depends only on $S$ and the $U^j$ 
and its minimization can give rise easily to the fixing of all these moduli. Furthermore since generic 
CY manifolds have of order of a hundred $U^j$ fields or more, and the fluxes $F_3,H_3$ have also a range
of possible quantized values, at the minimum there may be accidental cancelations
such that  there is a very tiny value for $<W_{flux}>=W_0$. Such tiny values would be needed to understand the
smallnes of the SUSY breaking scale compared to a large string scale $M_s$ not much below the Planck scale.

The above fluxes are unable to fix the values of the Kahler moduli. However in specific 
compactifications there are non-perturbative effects which induce superpotential terms involving 
the Kahler moduli. Examples of such non-perturbative effects are instanton effects induced by 
euclidean $D3$ instantons and gaugino condensation on $D7$-branes wrapping appropriate 
4-cycles in the CY. Such effects have typically an exponentially suppressed behavior of the form
$W{np}\simeq \sum_a exp(-B_aT_a)$ for some constants $B_a$. These effects combined with those 
induced by fluxes $W_{flux}$ have the potential to fix all the moduli of specific CY orientifold 
compactifications \cite{kklt,fernando}. Although a detailed example  with all the required properties, including a
realistic model and non-vanishing (but very small) cosmological constant, is still lacking,
it seems very likely that those ingredients have the potential to fix all moduli.

\begin{figure}[htbp]
\begin{center}
\includegraphics[angle=00,width=6cm]{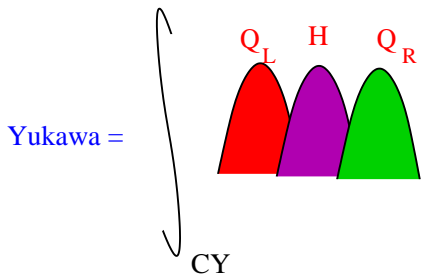}
\end{center}
\caption{ Pictorial representation of the computation of Yukawa coupling constants  as
overlap integrals of zero modes}
\label{overlapillo}
\end{figure}

Of more direct phenomenological interest are the Yukawa couplings involving SM quarks and leptons to
Higgs scalars. As we already mentioned, Yukawa couplings among $Dp$ brane matter fields in Type IIB
compactification arise from overlap integral of the wave function in extra dimensions of the three participant 
fields. Consider the case of D9-branes to simplify the discussion (recall the case of D7-branes may be described
in terms of D9-branes with appropriate fluxes). Suppose we have initially a $D=10$ Type IIB  orientifold with 
D9 branes and a gauge group $U(n)$. In the field theory limit our action will be 10d super-Yang-Mills,
\begin{eqnarray}
 {L} = -{1 \over {4}}  Tr\left(F^{MN}F_{MN}\right) +
{i \over {2}}  Tr\left(\bar{\Psi}\Gamma^M D_M \Psi\right) \, .
\label{10d-sym-yukawas}
\end{eqnarray}
We then compactify the theory on some CY manifold and turn on magnetic fluxes which
may break the gauge group to a SM-like gauge group. The 10d fields can then be expanded
a la Kaluza-Klein (KK)
\begin{equation}
\Psi (x^\mu, y^m) =  \sum_k  { \chi_{(k)} (x^\mu)} \otimes
 {\psi_{(k)}(y^m)} \, ,
A_{n}(x^\mu,y^m) =  \sum_k   {\varphi_{(k)}(x^\mu) } \otimes
 {\phi_{(k),n}(y^m)} \nonumber
\end{equation}
where $x^\mu$ and $y^m$ are 4d and internal coordinates respectively. The 4d massless spectrum 
may be chiral and $N=1$ supersymmetric with judicious choice of magnetic fluxes.
The 4d Yukawa coupling between these matter fields
arise from KK reduction of the cubic coupling $A\times \Psi \times\Psi$ from the 10d Lagrangian in eq.(\ref{10d-sym-yukawas}).
As illustrated in fig.(\ref{overlapillo}) the Yukawa coupling coefficients  are obtained from the overlap integrals
\begin{equation}
{Y_{ijk}}\,\,{=}\
{g \over 2} { \int_{CY}} {\small  {\psi_i^{\alpha \dag}}\,
\Gamma^m   \, {\psi_j^{\beta}}  { \phi_{k\,m}^{\gamma}} } 
f_{\alpha \beta \gamma} \, ,
\end{equation}
where $g$ is the 10d gauge coupling, $\alpha$, $\beta$, $\gamma$ are $U(n)$ gauge indices and $f_{\alpha \beta \gamma}$ are $U(n)$ structure 
constants; also $\psi$ , $\phi$ are fermionic and bosonic zero modes respectively, and $i,j,k$ label the different zero modes in a given charge sector, i.e. the families in semi-realistic models. The Yukawa couplings are thus obtained as overlap integrals of the three zero mode wave functions in the CY.

In order to compute the Yukawa coupling constants we thus need to know the explicit form of the
wave functions on compact dimensions of the involved matter fields, quarks, leptons and Higgs multiplets in a
realistic model.  However such wave functions are only accessible to explicit computation for simple
models like toroidal compactifications or orbifolds thereof. Indeed this computation has been worked out for
general toroidal/orbifold  models \cite{magneticus}. The wave functions turn out to be 
proportional to Jacobi  ${\theta}$-functions with a Gaussian profile and the holomorphic Yukawa couplings turn out to be
also proportional to products of  Jacobi $\theta$-functions (one per $T^2$ factor) with dependence only on
the complex structure moduli $U^j$ and the open string moduli (Wilson line degrees of freedom). 
As an example, the semirealistic model  in table (\ref{tabliguay}) has holomorphic Yukawa couplings with the structure
\begin{eqnarray}
\begin{array}{c}
 {Y_{ij}}^{ U} \sim
\vartheta \left[
\begin{array}{c}
\!\! \frac { i}{3}  \!\! \\ \!\!  0\!\!
\end{array}
\right] \left({3  {J^{(2)}} }\right)
\times
\vartheta \left[
\begin{array}{c}
\!\! \frac { j}3 \!\! \\
\!\!  0 \!\!
\end{array}
\right] \left({3  {J^{(3)}} }\right), \\[12pt]
 {Y_{ij*}}^{ D} \sim
\vartheta \left[
\begin{array}{c}
\!\!\frac { i}3  \!\!\\ \!\!
0  
\end{array}
\right] \left({3  {J^{(2)}}}\right)
\times
\vartheta  \left[
\begin{array}{c}
\!\!\frac { {j*}}{3}  \!\!\\ \!\!
 0 \!\!
\end{array}
\right] \left({3 {J^{(3)}} }\right),
\end{array}
\label{yukmatrices}
\end{eqnarray}
where $\vartheta$ is a Jacobi $\theta$-function, $J^{(i)}$ are the Kahler forms of the i-th torus and $i,j,j^*$ are family indices for the  $Q,U$ and $D$  SM chiral multiplets 
respectively. These expressions yield proportional expressions for $U$- and $D$-quark  Yukawa couplings  but they differ if one 
takes into account the generic possibility (in tori) of Wilson line backgrounds along the $T^6$ circles. However, due to the factorized 
structure of the family dependence, only one quark/lepton generation gets a mass.  The corresponding Yukawa coupling is 
 of order of the
gauge coupling constant. This may be considered as a good first approximation to the observed quark/lepton mass spectrum and 
one expects further corrections to give rise to the Yukawa couplings of the lighter generations. 

The computation of Yukawa couplings in general curved CY manifolds is more difficult, although it becomes more tractable
within the context of the  {\it bottom-up} approach mentioned above. The idea is that in models in which the SM fields are 
localized in brane  intersections, Yukawa couplings appear at points in the CY  in which three such intersections (corresponding to
SM and Higgs fields) meet. We already saw that  in the F-theory context in fig.\ref{unfolding}. Thus e.g. the Yukawa coupling
${\bf 10}\times {\bf {\bar 5}}\times {\bf {\bar 5}_H}$ in a $SU(5)$ F-theory GUT is localized at a point of triple intersection 
of the three matter curves.  The Yukawa coupling has now the schematic form
$\int_S \psi_i\psi_j\phi_H$ in which the integral, extended over the 4-fold $S$, is dominated by the intersection region.
In such a situation, to compute the Yukawa coupling we only need to know the wave functions  {\it in the  neighborhood }
of the intersection point \cite{yukftheory}. Those local wave functions may be obtained by solving the Dirac and K-G equations at the local
level. Interestingly, one again finds that  only one  (the third)  generation gets a  non-vanishing Yukawa coupling,
which is also of order the gauge coupling constant.  It has been found however that instanton corrections induced by
distant 7-branes  wrapping other 4-cycles in compact space in general induce the required Yukawa couplings
for the lighter generations \cite{mm,correcyuk}.

Instanton effects do not only give rise to superpotentials for the Kahler moduli and induce the Yukawa couplings of the 
lighter generations. They may also give rise to interesting terms in the SM superpotential which are forbidden in
perturbation theory. In particular, in brane models of SM physics  there are typically extra $U(1)$ gauged symmetries 
beyond those of the SM. A classical example is $U(1)_{B-L}$ which often appears gauged in many string constructions
including right-handed neutrinos. This symmetry is anomaly-free but there are often in addition $U(1)$'s with triangle anomalies
which are cancelled by the 4d version of the Green-Schwarz mechanism.  All anomalous $U(1)$'s become massive by
combining with the imaginary part of the Kahler(complex structure) moduli in Type IIB(IIA) orientifolds.  But,  in addition,  
anomaly-free gauge symmetries like $U(1)_{B-L}$ may  also become massive in this way. This happens due to the fact that 
e.g. in Type IIB some Im$T_i$ transform under the corresponding gauge $U(1)_a$ symmetries as Im$T_i\rightarrow$Im$T_i+q_i^a\Lambda_a$,
with $\Lambda_a$ the gauge parameter. This has interesting consequences for instanton physics \cite{instantons}.
 In a Type IIB orientifold some 
instanton configurations  corresponds to Euclidean D3-branes wrapping the compact dimensions (so that they are locallized in
Minkowski, as instantons should). If they intersect the  D7-branes where the SM fields live, there appear charged zero modes 
(from open string exchange) 
contributing to instanton induced transitions. This is why this class of stringy instantons are often called 
{\it charged} instantons.
 In particular if  the D3-brane wraps a 4-cycle with Kahler modulus $M$ (some linear combination 
of the $T_i$'s), non-perturbative operators of the general form
\begin{equation}
e^{-M}\Phi_{q_1}...\Phi_{q_n}  \ \ ,\ \ \sum_iq_i\not= 0
\end{equation}
may appear \cite{instantons}. These operators are gauge invariant because the sum of the charges of the
$\Phi$ chiral fields is compensated by the shift on Im$M$ induced by the gauge transformation. 
An example of this is the generation of right-handed neutrino masses in MSSM-like orientifolds
with a massive  $U(1)_{B-L}$ induced by a G-S mechanism. 
In this case the operator has the form $e^{-M}\nu_R\nu_R$ and the non-invariance of the 
bilinear under $U(1)_{B-L}$ is compensated by a shift of the Im$M$. The mass is of order 
$e^{-ReM}M_s$ which may be on the right  phenomenological ballpark 
$10^{12}-10^{14}$ GeV for $M_s\simeq 10^{16}$ GeV and $ReM\simeq 100$.
This type of {\it charged} instantons  could also be important for the generation of other
phenomenologically relevant terms 
like e.g. the  MSSM $\mu$-term.

\section{String model building and the LHC}

With the LHC in operation an important issue is trying to make contact between an underlying
string theory and experimental data. Of course it would be really exciting if the string scale 
$M_s$ was within reach of the LHC. We could perhaps observe some string or KK excitation
as resonances in LHC data. On the other a large string scale $M_s\simeq 10^{16}$ GeV seems to be 
favored if one sticks to a SUSY version of the SM such as the MSSM, in which gauge couplings nicely
unify at a scale of order $10^{16}$ GeV. 
So it is important to see whether specific classes of string
compactifications may lead to low energy predictions for SUSY breaking parameters.

We have seen that in certain large classes of Type II models there
is information about the structure of the low-energy effective action. In particular in Type IIB orientifolds
(or their F-theory extension) with SM fields localized at intersecting D7-branes (or matter curves in
F-theory GUT's) one can compute the  dependence on the local Kahler modulus of the 
gauge kinetic function (eq.(\ref{gaugekin})) and  also of the Kahler metric (eq.(\ref{metricswiss2})).
If a MSSM-like model is constructed on such a setting, one can  obtain specific expressions 
for SUSY breaking soft terms
assuming Kahler moduli dominance in SUSY breaking, i.e.,  non-vanishing auxiliary fields $F_t\not=0$.
This is a reasonable assumption within Type IIB/F-theory since in Type IIB orientifolds  such non-vanishing 
auxiliary fields correspond to the presence of non-vanishing antisymmetric RR and NS imaginary self-dual  $(0,3)$ fluxes \cite{camara},
which are known to solve the classical equations of motion \cite{gkp}. As we mentioned above, such closed string fluxes
are generically present in compactifications with fixed moduli. Using standard $N=1$ supergravity formulae and the 
above information on the effective action one obtains soft terms with the CMSSM structure but with the
additional  relationships \cite{aci1}, 
\begin{equation}
M\ =\ \sqrt{2}m\ = \ -(2/3)A\ =\ -B \, .
\label{semboundary}
\end{equation}
where $M$ is the universal gaugino mass, $m$ the universal scalar mass, $A$ the trilinear scalar parameter
and $B$ the Higgs bilinear parameter. Here one assumes the presence of an explicit $\mu$-term in the low energy
Lagrangian so that altogether there are only 2 free parameters,
$M$ and $\mu$. The universality of soft terms may be understood  if an underlying GUT structure exists as in
F-theory GUT's. As we have mentioned, magnetic flux backgrounds are generically present on the worldvolume
of the underlying 7-branes in order to get a chiral spectrum. In the presence of magnetic fluxes the gauge kinetic functions 
(see eq.(\ref{fluxcorrgauge})) and the Kahler metrics may get small corrections  to eqs.(\ref{gaugekin}),(\ref{metricswiss2}), i.e.
\begin{equation}
f\ =\ T( 1\ + \ a \frac{S}{T} )\ \ ;\ K_\alpha \ =\ \frac {t^{1/2}}{t_b} (1\ +\  \frac{c_\alpha}{t^{1/2} } )\  ,
\end{equation}
where $a$ and $c_\alpha$ are constants and $S$ is the  the complex dilaton field.
These  corrections are suppressed in the large $t$ limit, corresponding to the physical weak coupling.
In this limit one may also neglect the correction to $f$ compared to that coming from $K_\alpha$.
One then finds corrected soft terms of the form
\begin{eqnarray}
m_{\tilde f}^2 \  & = & \ \frac{1}{2}|M|^2(1-\frac {3}{2}\rho_f)\,, \\  
m_{H}^2 \ & = & \   \frac{1}{2}|M|^2  (1- \frac{3}{2}\rho_H)\,,  \\
A \ & = & \ -\frac{1}{2}M(3\ -\ \rho_H\ -\ 2\rho_f)\,, \\ 
B\ & = & \ -M(1-\rho_H)  \, ,
\label{boundconditionsfinal}
\end{eqnarray}
where $\rho_\alpha=c_\alpha/t^{1/2}$ and the subindices $f,H$ refer to fluxes through the
fermion matter curves or the Higgs curve.
Note that as an order of magnitude one numerically expects 
$\rho_{H,f}\simeq 1/t^{1/2}\simeq \alpha_{GUT}^{1/2}\simeq 0.2$. 
\begin{figure}[t!]
\hspace*{-0.6cm}
\begin{center}
\includegraphics[width=9.5cm, angle=270]{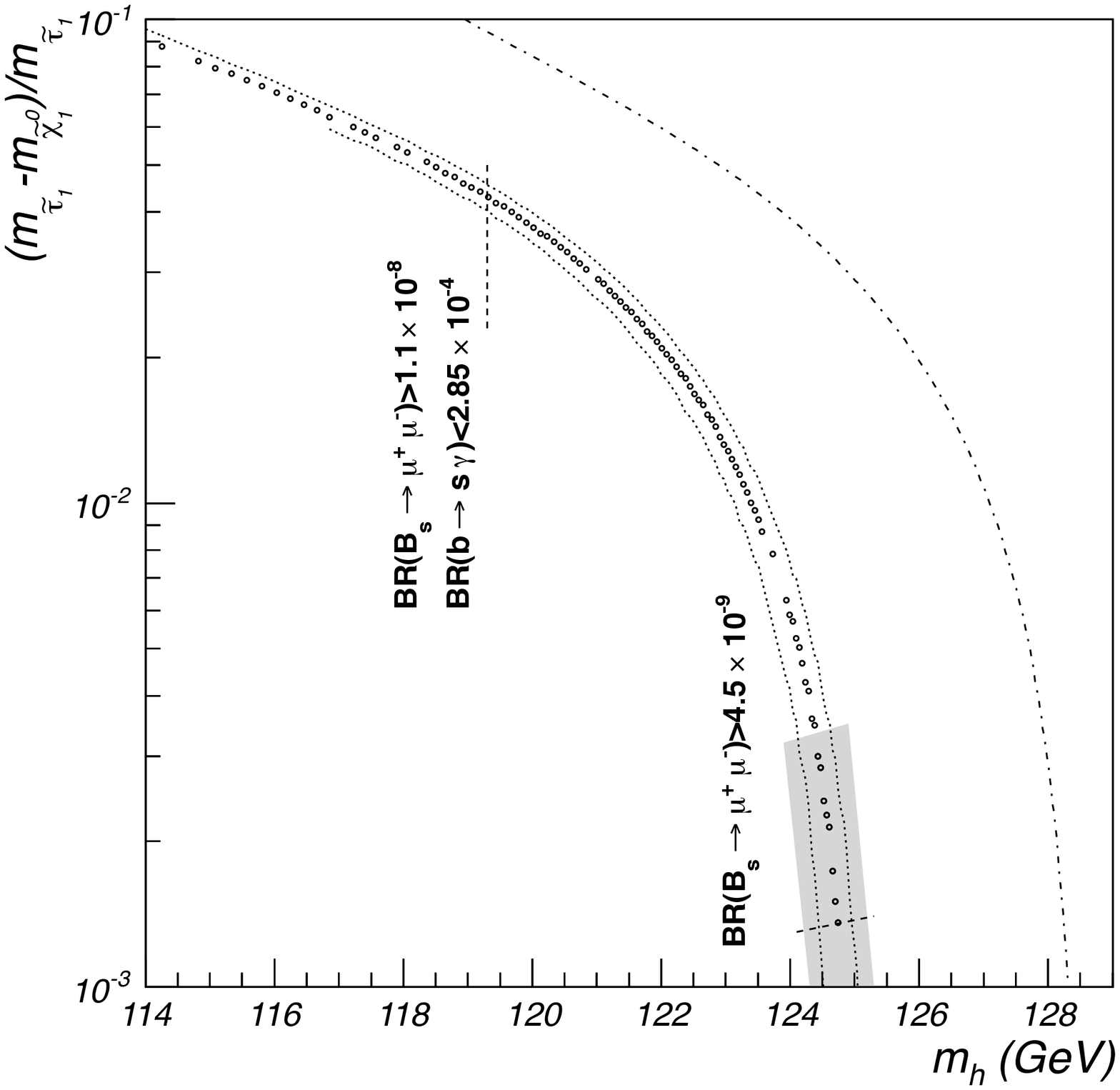}
\end{center}
\caption{The normalized mass difference 
$(m_{{\tilde \tau}_1}-m_{\chi_1^0})/m_{{\tilde \tau}_1}$ as a function of the lightest Higgs mass $m_h$ in the modulus dominance scheme.
Appropriate REWSB, neutralino dark matter and  BR$(B_s\to\mu^+\mu^-)$ limits are only consistent for a Higgs mass in the 125 GeV region.
(from ref.\cite{aci2}).}
\label{coannihilation}
\end{figure} 
The above soft terms apply at the string/unification scale $M_s\simeq 10^{16}$ GeV. 
In order to get the low energy physics around the EW scale one has to run down the 
soft parameters according to the renormalization group equations (RGE). 
Then one has to check that the boundary conditions are consistent with radiative EW symmetry breaking (REWSB)
and with present low-energy phenomenological constraints. One may in addition impose that the lightest
neutralino  is stable and provides for the dark matter in the universe.  The resulting scheme is extremely
constrained \cite{aci2}. In particular, setting the fermion flux correction to zero for simplicity, one has a theory with 
three free parameters ($M,\mu$ and $\rho_H$) and two constraints (REWSB and dark matter), or equivalently,
lines in the planes of any pair of parameters or SUSY masses. As an example fig.(\ref{coannihilation}) shows 
the normalized mass difference 
$(m_{{\tilde \tau}_1}-m_{\chi_1^0})/m_{{\tilde \tau}_1}$ as a function of the lightest Higgs mass $m_h$
\cite{aci2}.  Dots correspond to points fulfilling the central value in the result from WMAP for the neutralino relic density and dotted lines denote the upper and lower limits after including the $2\sigma$ uncertainty. 
The dot-dashed line represents points with a critical matter density $\Omega_{matter}=1$. The vertical line corresponds to the 2$\sigma$ limit on BR($b\rightarrow s\gamma$) and the upper bound on BR$(B_s\to\mu^+\mu^-)$ from Ref.\,\cite{cmslhcb} and the recent LHCb result \cite{Aaij:2012ac}. The gray area indicates the points compatible with the latter constraint when the $2\sigma$ error associated to the SM prediction is included.
As is obvious from the figure, the dark matter condition is fulfilled thanks to a stau-neutralino coannihilation mechanism.  Interestingly enough, the
recent constraint on BR$(B_s\to\mu^+\mu^-)$ from LHCb forces the Higgs mass to a region around 125 GeV, consistent with the hints of
a Higgs particle in that range as measured at CMS and ATLAS.
Fixing the mass of any SUSY particle fixes the rest of the spectrum. In particular, with a lightest Higgs mass around  125 GeV gluinos have a mass
around 3 TeV,  the 1-st,2-nd generation squarks around 2.7-2.8 TeV and the lightest stop
around 2 TeV.   The lightest slepton is a stau with mass around 600 GeV, almost degenerate
with the lightest neutralinos.  The existence of  gluino and squarks of these masses can be tested at LHC running at 14 TeV and 30 fb$^{-1}$
integrated luminosity.  

It is remarkable that a lightest MSSM Higgs mass as heavy as 125 GeV is possible in this scheme. 
In most SUSY schemes  (including minimal gauge and anomaly  mediation models and the CMSSM with not superheavy squarks) the lightest Higgs mass
is typically around 115 GeV or so \cite{higgsimpact}.
In this scheme a relatively heavy Higgss appears  because the 
soft terms in eq.(\ref{boundconditionsfinal}) predict a large $A$-parameter with $A\simeq -2m$, giving rise to a large stop
mixing parameter and hence a big one-loop correction to the Higgss mass. In addition the dark matter and REWSB conditions 
require a large tan$\beta\simeq 40$, pushing the tree level Higgss mass to its maximum value.  This large tan$\beta$ and 
stop mixing parameters imply that,
as it stands this simple scheme may be soon ruled out if LHCb finds  no deviation from the SM value for 
BR$(B_s\to\mu^+\mu^-)$, which may happen soon. On the other hand  a  NMSSM version of the same model,
also viable in Type IIB/F-theory schemes,  would remain consistent as would also R-parity violation, since it  would 
avoid the dark matter over-abundance problem.
This shows how the LHC results may provide 
important constraints on the possible  compactifications and SUSY-breaking schemes within string theory,
see e.g. ref.(\cite{othersoft}) for other string derived  approaches.

\vspace*{2cm}

\begin{center}{\bf Acknowledgments}\\\end{center}

I thank the organizers of the school  for their kind invitation. I am also grateful 
to L. Aparicio, D. G. Cerde\~no, A. Font, P.G. C\'amara, F. Marchesano, F. Quevedo, A. Uranga for many useful discussions and collaboration.
This work has been partially supported by the grants FPA 2009-09017, FPA 2009-07908, Consolider-CPAN (CSD2007-00042) from the MICINN, HEPHACOS-S2009/ESP1473 from the C.A. de Madrid and the contract ``UNILHC" PITN-GA-2009-237920 of the European Commission.

\newpage


\end{document}